\documentclass[12pt]{article}
\pdfoutput=1
\usepackage[utf8]{inputenc}
\usepackage[T1]{fontenc}
\usepackage{float}
\usepackage{authblk}
\usepackage[normalem]{ulem}
\usepackage{amsfonts}       
\usepackage{amsmath}
\usepackage{amsthm}
\usepackage{amsbsy}
\usepackage{amssymb}
\usepackage{nicefrac}   
\usepackage{centernot}

\usepackage{adjustbox}
\usepackage{graphicx}
\usepackage{booktabs}
\usepackage{mathtools}
\usepackage{fullpage}
\usepackage[mathlines]{lineno}
\usepackage{subcaption}
\usepackage[style=numeric-comp,sorting=none,maxnames=99]{biblatex} 

\makeatletter
\renewcommand*\env@matrix[1][\arraystretch]{%
    \edef\arraystretch{#1}%
    \hskip -\arraycolsep
    \let\@ifnextchar\new@ifnextchar
    \array{*\c@MaxMatrixCols c}}
\makeatother

\usepackage{mathtools}

\usepackage{multirow}
\usepackage[table,xcdraw]{xcolor}

\usepackage[colorlinks = true,
            linkcolor = teal,
            urlcolor  = teal,
            citecolor = teal]{hyperref}

\usepackage[font=footnotesize,labelfont=bf]{caption}

\title{Particular flows and attracting sets \\ A comment on "How particular is the physics of the Free Energy Principle?" by Aguilera, Millidge, Tschantz and Buckley}

\author[1-4]{Conor Heins\thanks{\href{mailto:cheins@ab.mpg.de}{cheins@ab.mpg.de}}}

\affil[1]{Department of Collective Behaviour, Max Planck Institute \protect\\of Animal Behavior, 78464 Konstanz, Germany}
\affil[2]{Department of Biology, University of Konstanz, 78464 Konstanz, Germany}
\affil[3]{Centre for the Advanced Study of Collective Behaviour, \protect\\University of Konstanz, 78464 Konstanz, Germany}
\affil[4]{VERSES Research Labs, Los Angeles, California, USA}

\begin{document}
\maketitle

\pagenumbering{arabic}


\begin{abstract}
    In this commentary, I expand on the analysis of the recent article "How particular is the physics of the Free Energy Principle?" by Aguilera et al. by studying the flow fields of linear diffusions, and particularly the rotation of their attracting sets in the presence of different types of solenoidal coupling. This analysis sheds new light on previous claims made in the FEP literature (and contested in the target article) that the internal dynamics of stochastic systems can be cast performing a gradient flow on variational free energy, and thus endowed with an inferential interpretation, i.e., as if internal states are performing inference about states external to the system. I express general agreement with the target article's statement that the marginal flow of internal states does not point along variational free energy gradients evaluated at the most likely internal state (i.e., the conditional mode). However, in this commentary I focus on the flow of particular states (internal and blanket states) and their variational free energy gradients, and show that for a wide but restricted class of solenoidal couplings, the average flow of these systems do indeed point along variational free energy gradients. This licenses a different but perhaps stronger re-description of the flow of particular states as performing inference, which importantly holds at arbitrary points in state space, not just at the conditional modes.
\end{abstract}

\section{Introduction}

The target article, "How particular is the physics of the Free Energy Principle" \cite{aguilera2021particular}, investigates the Free Energy Principle (hereafter: FEP) in the context of linear diffusion processes, i.e., Ornstein-Uhlenbeck processes \cite{pavliotisStochasticProcessesApplications2014,godrecheCharacterisingNonequilibriumStationary2019}. Using this class of systems as a testbed for the FEP's claims, the authors carefully deconstruct various assumptions and their consequences, which they claim are made (implicitly or explicitly) in the existing FEP literature, specifically in works such as \cite{friston2019free, parr2020markov, friston2021some}. In this commentary, I offer an additional perspective on the claims of the target article regarding the marginal flows of the conditional modes  \cite[Section 3.2]{aguilera2021particular}, which afford the inferential interpretation of the FEP described in \cite{friston2019free, parr2020markov}. Here I study system-level flows by focusing on the average flow of so-called `particular states', and specifically analyze the conditions under which they license an inferential interpretation of the FEP, particularly the case when particular states can be described as flowing along the gradients of the variatonal free energy.

Consider a system whose state $x$ is composed of internal $\mu$, blanket $b$, and external $\eta$ states $x=(\eta,b, \mu)$ and whose evolution is described by an Itô stochastic differential equation
\begin{align}
\label{eq: sparsely coupled SDE}
\begin{bmatrix}
\dot \eta\\
\dot b\\
\dot \mu
\end{bmatrix} = 
\begin{bmatrix}
f_\eta(x)\\
f_b(x)\\
f_\mu(x)
\end{bmatrix}+ 
\begin{bmatrix}
\omega_\eta\\
\omega_b\\
\omega_\mu
\end{bmatrix},
\end{align}
where $\omega_\eta, \omega_b, \omega_\mu$ are independent Wiener processes.
Let us assume the existence of a steady-state distribution $p$ such that $b$ is a Markov blanket between external and internal states
\begin{align*}
   \eta \perp \mu \mid b \iff p(\mu, \eta|b) = p(\mu | b) p(\eta | b).
\end{align*}
Henceforth, the \emph{conditional modes} refer to the most likely internal (respectively external) states, conditioned on blanket states, assuming they exist \cite{da2021bayesian}
\begin{align}
\label{eq: most likely internal external}
    \boldsymbol \mu (b)\triangleq \arg\max p(\mu \mid b), \quad \boldsymbol \eta (b)\triangleq \arg\max p(\eta \mid b).
\end{align}





The \textit{marginal flow} of the system given a blanket state refers to the flow averaged over external and internal states given blanket states \cite[Equation 15]{aguilera2021particular}
\begin{align}
     \dot{\boldsymbol \mu} (b) \triangleq \mathbb{E}_{p(\eta, \mu | b)}[f_\mu (\mu, b, \eta)], &\quad \dot{\boldsymbol \eta} (b) \triangleq\mathbb{E}_{p(\eta, \mu | b)}[f_\eta (\mu, b, \eta)]. \label{eq:marginal_flow_def}
\end{align}

The authors analyze the marginal flows of linear systems and contest what they claim is a central tenet of the FEP: namely, that the marginal flows of internal states $\dot{\boldsymbol \mu}$ point down the gradients of the free energy $F$ \cite[Equation 22]{aguilera2021particular},
\begin{align}
\label{eq: alignment claim}
    \dot{\boldsymbol \mu} (b) &\propto -\nabla_{\boldsymbol \mu(b)}F(\boldsymbol \mu(b), b),
\end{align}
where the variational free energy $F$ allows one to relate the dynamics of random systems with approximate Bayesian inference in ~\cite{friston2019free, parr2020markov}.




Here I will expand on the target article's  discussion about the relationship between variational free energy gradients and the marginal flow of internal states $\dot{ \boldsymbol \mu} (b)$ in linear systems, offering a slightly different perspective in terms of particular states. Specifically, I use the flows of particular states $f_{\mu, b}(\mu, b, \eta)$ to demonstrate the space of flows that operate on particular states --- including, importantly, the flow at the conditional mode $\boldsymbol \mu (b)$. This analysis reveals a complex relationship between solenoidal flow, the particular flows and the variational free energy gradients. 
This analysis also finds agreement with target article's overarching result that the dynamics of the conditional modes are not representative of system-wide behaviour\footnote{However, note that this does not conflict with the central claims of the FEP.}.

However, these results also supplement the discussion about marginal flows, and suggest that the relationship between solenoidal flows and the gradient flows of internal states is more nuanced than either the target article or previous FEP literature suggests. I show that the flow of \textit{particular states} $f_{\mu, b}(\mu, b, \eta)$ under different external conditions is always pulled to a linear attracting set, but importantly, this attracting set is only aligned with the line of conditional modes $\boldsymbol \mu (b)$ in special cases. These cases depend on the nature of solenoidal couplings. Thus, the flow of internal states (and thus the marginal flow)
does not generally point along variational free energy gradients, consistent with the demonstrations of the target article \cite{aguilera2021particular}. I thus express general agreement with the target article in questioning the generality of statements made in earlier FEP literature (e.g., \cite[Equation 8.26]{friston2019free} and \cite[Equation 3.3]{parr2020markov}).

However, the conditions under which internal states flow along free energy gradients are also not as restrictive as the target article suggests. Indeed, I show in the context of simple linear diffusions with a Markov blanket between internal and external states that the presence of solenoidal couplings do not \textit{always} misalign the attracting sets of particular states with the line of the conditional mode $\boldsymbol \mu (b)$. This implies that the flow of the particular states (which crucially includes the internal states and their conditional mode) can still point along variational free energy gradients even in the presence of solenoidal coupling. Indeed, our results suggest that the attracting set of particular states is identical to the line of the conditional modes for a large class of solenoidal couplings. For instance, I show that solenoidal coupling between external and blanket states rotates the linear attracting set of particular states relative to the line of the conditional modes, but in doing so also contracts the effective volume of state space that particular states visit. This solenoidal coupling helps particular states accelerate their flow towards (and maintain their proximity to) the free energy minimum. 
This is in line with investigations into the importance of solenoidal coupling in augmenting the speed of convergence of diffusions to their steady-state \cite{hwangAcceleratingDiffusions2005, guillin2016optimal, barpGeometricMethodsSampling2022a,lelievreOptimalNonreversibleLinear2013}, and hints at the importance of the `canonical' or `normal' form of the flow introduced in previous FEP literature \cite{friston2019free,friston2021some, friston2021stochastic, friston2022free}.

\section{Solenoidal rotation of attracting sets}

Recall the 3-dimensional linear diffusion described in Equation \eqref{eq: sparsely coupled SDE}, whose state $x$ is composed of internal $\mu$, blanket $b$, and external $\eta$ states, i.e., $x = (\eta, b,\mu)$. The system's stochastic evolution is described using the following (Itô) stochastic differential equation
\begin{align}
    \begin{bmatrix} \dot \mu \\ \dot b \\ \dot \eta \end{bmatrix} &= \underbrace{\begin{bmatrix} \mathbf{J}_{\mu \mu} & \mathbf{J}_{\mu b} & \mathbf{J}_{\mu \eta} \\ \mathbf{J}_{b \mu} & \mathbf{J}_{b b} & \mathbf{J}_{b \eta} \\ \mathbf{J}_{\eta \mu} & \mathbf{J}_{\eta b} & \mathbf{J}_{\eta \eta}\end{bmatrix} \begin{bmatrix} \mu \\ b \\ \eta \end{bmatrix}}_{=f(x)} + \begin{bmatrix}
\omega_\eta\\
\omega_b\\
\omega_\mu
\end{bmatrix}, \label{eq:Ito_form}
\end{align}

where $\mathbf{J}$ is the Jacobian of the flow $f$ in Equation \eqref{eq: sparsely coupled SDE}. Assuming that the spectrum of the Jacobian has negative real parts, we have the existence of a Gaussian (non-equilibrium) steady-state $p(x) = \mathcal N(x;0, \boldsymbol \Pi^{-1})$ with precision matrix $\boldsymbol \Pi$ \cite{godrecheCharacterisingNonequilibriumStationary2019}. Thus, the flow can be rewritten via the Helmholtz decomposition \cite[Appendix B]{da2021bayesian} as
\begin{align}
    \mathbf{J} x = -(\boldsymbol \Gamma + \boldsymbol Q)\nabla_x \ln p(x), \label{eq:HH_decomp}
\end{align}
into a dissipative, curl-free component $\boldsymbol \Gamma \nabla_x \ln p(x)$ and a conservative, divergence-free component $\boldsymbol Q \nabla_x \ln p(x)$, where each component depends on the gradients of the stationary density $p(x)$. The diffusion tensor $\boldsymbol \Gamma$ is a positive-definite matrix that determines the amplitude and covariance of random fluctuations $\boldsymbol \Gamma \triangleq \mathbb E[\omega \omega^\top]/2$. The skew-symmetric matrix $\boldsymbol Q=-\boldsymbol Q^\top$ mediates conservative, rotational dynamics in the system's flow. 

I begin by analyzing linear systems that allow for non-zero solenoidal coupling between internal and blanket states, and between blanket states and external states. As in the earlier FEP literature \cite{friston2019free, parr2020markov, friston2021some, biehl2021technical, friston2021stochastic, friston2022free}, and consistent with one of the `canonical flow constraints' also entertained in the target article and mentioned in a FEP literature \cite[12.11]{friston2019free}, I assume there is no solenoidal coupling between internal and external states $\mathbf{Q}_{\eta \mu} = -\mathbf{Q}_{\mu \eta}^\top = 0$. Given these constraints, one can parameterize the solenoidal matrix $\boldsymbol Q$ as follows
\begin{align}
    \boldsymbol Q (\gamma, \lambda) &= \begin{bmatrix} 0 & \gamma & 0 \\ -\gamma & 0 & -\lambda \\ 0 & \lambda & 0\end{bmatrix} = \begin{bmatrix} 0 & \mathbf{Q}_{\eta b} & 0 \\ -\mathbf{Q}_{\eta b} & 0 & \mathbf{Q}_{b \mu} \\ 0 & -\mathbf{Q}_{b \mu} & 0\end{bmatrix}.
\end{align}

Under this simple parameterisation, $\gamma$ determines the strength of solenoidal coupling between $b$ and $\eta$, and $\lambda$ the strength of solenoidal coupling between $b$ and $\mu$.
For all numerical demonstrations, I fix the inverse of the stationary covariance---also known as the precision matrix $\boldsymbol \Pi$---and the diffusion tensor $\boldsymbol \Gamma$ to the following values
\begin{align}
    \boldsymbol\Sigma^{-1} \triangleq \boldsymbol \Pi &= \begin{bmatrix}2 & 1 & 0 \\ 1 & 2 & 1 \\ 0 & 1 & 2 \end{bmatrix}, \quad \boldsymbol \Gamma = \begin{bmatrix} 2 & 0 & 0 \\ 0 & 2 & 0 \\ 0 & 0 & 2 \end{bmatrix}.
\end{align}
Note that $\boldsymbol \Pi_{\mu \eta} = \boldsymbol \Pi_{\eta \mu} = 0$ is equivalent to the existence of a Markov blanket between internal and external states \cite[Equation 2.2]{da2021bayesian} (note that this holds independently of solenoidal flow).

In the case of Gaussian distributions, the product of the precision matrix with the state of the system $x$ is a vector whose entries contain the gradients of the log probability density $\boldsymbol\Pi x = \nabla_x \ln p(x)$. This affords a simpler expression for the flow $f(x)$
\begin{align}
    f(x) = \mathbf{J}x  &= \underbrace{\begin{bmatrix} -2 & -\gamma & 0 \\ \gamma & -2 & \lambda \\ 0 & -\lambda & -2 \end{bmatrix}}_{=-(\boldsymbol \Gamma + \boldsymbol Q)} \underbrace{\begin{bmatrix}2 & 1 & 0 \\ 1 & 2 & 1 \\ 0 & 1 & 2 \end{bmatrix} \begin{bmatrix}\mu \\ b \\ \eta \end{bmatrix}}_{=\nabla_x \ln p(x)}
\end{align}

\begin{figure}[!ht]
    \centering
    \includegraphics[width=0.75\textwidth]{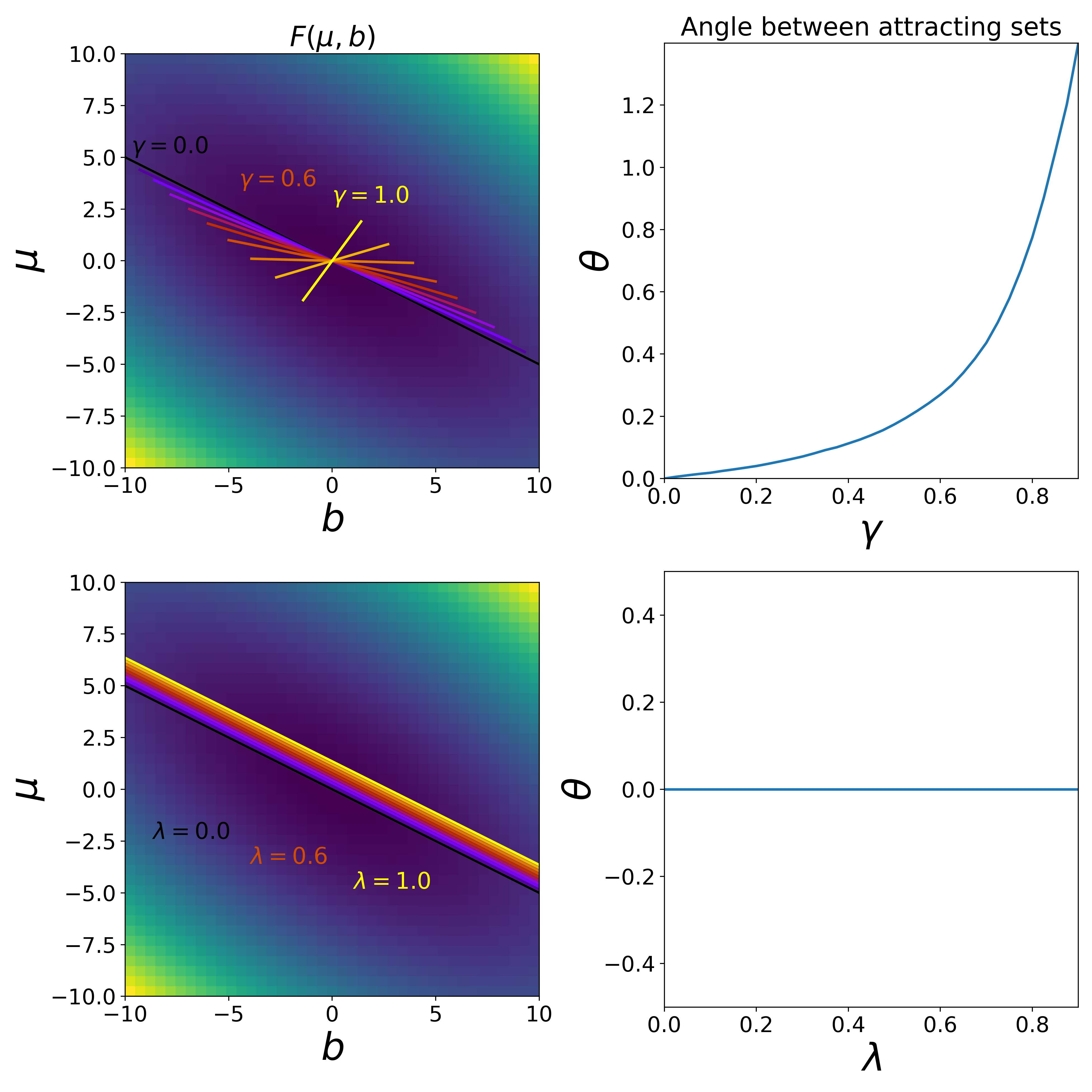}
    \caption{Relationship between solenoidal coupling and the rotation of attracting sets. \emph{Top row:} Solenoidal coupling between external and blanket states $\gamma = Q_{\eta b} = -Q_{b \eta}^\top$ rotates the attracting set of particular states away from the line of the conditional mode $\boldsymbol \mu (b)$. The left panel shows the attracting set of the particular flow $f_{\mu , b}(x)$, computed by evaluating the flow of $\mu, b$ over a fixed range of external states $\eta \in [-15, +15]$, for increasing values of solenoidal coupling strength $\gamma$, with $\lambda = -Q_{b \mu}$ fixed to $1.0$. The attracting sets are plotted on top of the variational free energy of particular states $F(\mu, b)$. The free energy attains its minimum along the line $(b,\boldsymbol \mu (b))$, meaning the flow of the internal states points towards the conditional mode (and hence down free energy gradients) when $\gamma = 0$. The right panel shows the angle (in radians) between the attracting set of $f_{\mu, b}( \mu, b, \eta)$ and the line of $\boldsymbol \mu (b)$ for increasing solenoidal coupling strength $\gamma$. The attracting set of the particular states is defined as the solution set of $\mu, b$ at which the flow of particular states $f_{\mu, b}(\mu, b, \eta)$ vanishes.
    \emph{Bottom row:} identical analysis as the top row but with varying the solenoidal coupling between internal and blanket states $\lambda= -Q_{b \mu} = Q_{\mu b}$. Here, solenoidal coupling between external and blanket states $\gamma = Q_{\eta b}$ was fixed to $0.0$. The attracting set of particular states does not change as a function of $\lambda$, but they are shown vertically offset from one another for visibility. The alignment between the attracting set and the line of the conditional mode persists despite changes in solenoidal coupling, demonstrating how such dynamics do not preclude an inferential interpretation of such systems.}
    \label{fig:1}
\end{figure}

Note that this construction means that although there is a Markov blanket between internal and external states (because of $\boldsymbol \Pi_{\mu \eta} = \boldsymbol \Pi_{\eta \mu} = 0$), in the presence of any solenoidal coupling (either $\gamma$ or $\lambda$ is non-zero), then there will be at least unidirectional coupling between internal and external states ($\mathbf{J}_{\eta \mu} \neq 0$ or $\mathbf{J}_{\mu \eta} \neq 0$, or both). Note that according to recent work \cite{friston2021stochastic, friston2022free}, `sparse coupling' entails either absent \textit{or unidirectional coupling}, so a system where $\mathbf{J}_{\eta \mu} = 0, \mathbf{J}_{\eta \mu} \neq 0$ still technically satisfies sparse coupling. 

In Figure \ref{fig:1} I explore how $\gamma$ and $\lambda$ determine the attracting set of the particular flow, defined as
\begin{align*}
    \{(\mu, b) \text{ s.t. } f_{\mu, b}(x)=0 \text{ for some external state } \eta\}.
\end{align*}
In the top row of Figure \ref{fig:1}, one can see that particular states are attracted to a line that is misaligned with the line $\{(b, \boldsymbol \mu (b))\}$---the set at which the free energy is minimised \cite{da2021bayesian}. In particular, the angle between the attracting set of particular states and the line $\{(b, \boldsymbol \mu (b))\}$ grows with increasing $\gamma$. When $\gamma = 0$, both are aligned and the flows of $\mu, b$ are attracted towards $\{(b, \boldsymbol \mu (b))\}$. However, as soon as $\gamma \neq 0$, the particular states are pulled towards an attracting set that is misaligned with $\{(b, \boldsymbol \mu (b))\}$ and thus does not point along variational free energy gradients, i.e.,
\begin{align}
    f_{\mu}(\mu, b, \eta) &\not \propto -\nabla_{\boldsymbol \mu (b)}F(\boldsymbol \mu(b), b) \notag \\
    &\not \propto -\nabla_{\mu}F(\mu, b).
\end{align}

Note that in this expression I include both the gradients of the free energy evaluated at the conditional mode $F(\boldsymbol \mu(b), b)$ as well as that defined over the space of particular states $F(\mu, b)$. Although statements concerning the gradients of the second quantity $F(\mu, b)$ are not central to the claims of the target article or the previous FEP literature, I include it as an interesting case to examine, because as we will see below, there indeed are parameterizations of systems where such gradients \emph{do} align with the flow of particular states. I also remark that the gradients of the variational free energy evaluated at the conditional mode, with respect to the conditional mode $\nabla_{\boldsymbol \mu} F(\boldsymbol \mu, b)$ are trivially $0$, so that quantity isn't of particular interest when we are trying to compare it to the marginal flows of internal states. Assuming any non-zero marginal flows in different parts of state-space, then the equality of these marginal flows with the variational free energy gradients (evaluated at the conditional mode, and with respect to the conditional mode) will trivially not hold. This is why expanding our analysis to the free energy gradients of the particular states, with respect to arbitrary internal states (not just the most likely ones) $\nabla_{\mu} F(\mu, b)$ is both a stronger and more interesting interpretation, because those gradients can be non-zero.

Alignment of the flow of internal states (including that of the conditional mode $\boldsymbol \mu (b)$) with the line $\{(b, \boldsymbol \mu (b))\}$ does not require an absence of solenoidal coupling. In the current example, all that is required for alignment is the absence of certain solenoidal couplings, in particular that between external and blanket states $Q_{\eta b}=0$. In general, arbitrary solenoidal coupling between internal and blanket states will not `misalign' the attracting set with the line $\{(b, \boldsymbol \mu (b))\}$, provided there is no solenoidal coupling between external and blanket states. That means that in such systems, internal and blanket states \textit{do} flow along the gradients of the variational free energy $F(\mu, b)$, towards their attracting set which lies along $(b, \boldsymbol \mu (b))$.

It is clear from the top row of Figure \ref{fig:1} that $\gamma$ rotates the attracting set of particular states. Interestingly, this rotation goes hand-in-hand with a simultaneous contraction of the attracting set's effective volume, which is evident from inspecting the change in length of the colored lines in Figure \ref{fig:1}, which span the attracting points of $f_{\mu, b}(x)$ evaluated over the same range of external states. The bottom row of Figure \ref{fig:1} shows the same attracting sets while varying the solenoidal coupling between blanket and internal states. In this case, the attracting set of particular states remains aligned with the line of conditional modes, and the volume of this attracting set does not change. This alignment persists for all strengths of the coupling parameter $\lambda$ between internal and blanket states, provided solenoidal coupling between external and blanket states is constrained to be absent (i.e. $\gamma = 0$).

These simple demonstrations show how solenoidal coupling relates to the alignment of flow fields with the variational free energy gradients and the line of the conditional modes (the point at which the variational free energy gradients vanish). For instance, our results call into question the claim of the target article that only systems with block diagonal solenoidal coupling sustain marginal flows of internal states that minimize free energy. Indeed, in the linear diffusions I examined here, the only constraint that is needed for the attracting sets to align and hence for internal states to minimize free energy is an absence of solenoidal coupling between external and blanket states ($\gamma = 0$). In the other cases where $\lambda$ is unconstrained, the particular flows find their conditional minima along the line of the conditional modes for different settings of $\eta$, and hence point along variational free energy gradients. 

I suspect (and this is intimated in more recent FEP literature such as \cite{friston2021stochastic,friston2022free}) that this result generalises to the setting of particular partitions, where the state space is further partitioned into autonomous states (internal and active) and non-autonomous states (external and sensory). In general, I speculate that such alignment would persist in systems where solenoidal coupling is absent between autonomous and non-autonomous states, but unconstrained \textit{within} either subset.



\section*{Additional information}\label{sec:additional_info}

\subsection*{Acknowledgements}\label{sec:ack}
I would like to express many thanks to Lancelot Da Costa for helpful discussions and edits that substantially contributed to the notation, presentation and overall quality of the commentary. I would also like to thank Thomas Parr and Karl Friston for helpful feedback and discussions that improved the commentary.

\subsection*{Funding statement}\label{sec:fund}
CH is supported by the U.S. Office of Naval Research (N00014-19-1-2556), and acknowledges the support of a grant from the John Templeton Foundation (61780). The opinions expressed in this publication are those of the author(s) and do not necessarily reflect the views of the John Templeton Foundation.

\printbibliography[title={References}]

\end{document}